\documentclass[a4paper,twoside]{article}      
\usepackage{amsmath,amssymb,amsfonts,amsthm,amscd} 
\usepackage{graphics}                 
\usepackage{color}                    
\usepackage{hyperref}                
\usepackage{ mathrsfs }
\usepackage{indentfirst}
\usepackage{bbold}
\usepackage{enumerate}
\usepackage{url}         
\usepackage{colonequals} 
\usepackage{a4wide}

\hypersetup{
    colorlinks=true,
    pdfborder={0 0 0},
}                               

\newtheorem{theorem}{Theorem}[section]

\theoremstyle{definition}
\newtheorem{remark}[theorem]{Remark}

\def\N{\mathbb{N}}

\newlength{\boxwidth}
\title{Weakly Non-Equilibrium Properties of Symmetric Inclusion Process with Open Boundaries}
\author{Kiamars Vafayi and Manh Hong Duong\\
\small Department of Mathematics and Computer Science,\\
\small Technische Universiteit Eindhoven,\\
\small Postbus 513, 5600 MB Eindhoven,\\
\small The Netherlands \\
\small \texttt{k.vafayi@tue.nl} and \texttt{m.h.duong@tue.nl}}

\date {\today}

\begin{document}

\maketitle

\section*{Abstract}
We study close to equilibrium properties of the one-dimensional Symmetric Inclusion Process (SIP) by coupling it to two particle-reservoirs at the two boundaries with slightly different chemical potentials. The boundaries introduce irreversibility and induce a weak particle current in the system. We calculate the McLennan ensemble for SIP, which corresponds to the entropy production and the first order non-equilibrium correction for the stationary state. We find that the first order correction is a product measure, and is consistent with the local equilibrium measure corresponding to the steady state density profile.

\section{Introduction}
A central concept in equilibrium statistical mechanics is the Gibbs-Boltzmann ensemble~\cite{Gibbs}
\begin{equation*}
 \rho(x)\propto e^{-\beta U(x)},
\end{equation*}
which relates the probability $\rho(x)$ of finding a system in a state $x$ at an inverse temperature of $\beta$ to its energy function $U(x)$. 
Out of equilibrium, however, the situation is more complex and there is no simple result analogous of the Gibbs-Boltzmann distribution. There have been many attempts to provide a general formalism for the non-equilibrium statistical mechanics, see e.g. \cite{bertini, Bertini09} and the references therein.

Microscopic models, such as stochastic lattice gases or interacting particle systems~\cite{Liggett85, Liggett99} and interacting diffusions~\cite{Varadhan79}, have been found to be useful in understanding non-equilibrium phenomena.
One big class of models are those of particle and heat transport, being used to model very diverse phenomena, from phase transitions to condensation and heat and mass transport.
To be able to tackle the non-equilibrium problem, instead of considering the general non-equilibrium situation, one approach is to study systems in contact to two particles or heat reservoirs at the boundaries, the so-called boundary driven systems \cite{Eyink}. These, in a sense, constitute the simplest and most controllable non-equilibrium settings. 
In this realm, exactly solvable models play an important role, as they enable us to test various ideas and concepts about non-equilibrium~\cite{schutz}.

A concept about non-equilibrium systems is the \emph{local equilibrium}. Intuitively, it says that although on a macroscopic level the thermodynamic variables might vary significantly in space, there are regions of smaller scale which have nearly constant macro variables. Therefore, we might approximately consider such regions in equilibrium. It is important to understand the criteria and situations in which the local equilibrium holds, for example in research related to hydrodynamic limit of different microscopic systems~\cite{sasa}. This of course can be studied in two settings, for instance in thermodynamic limit where the size of the system becomes very large. Or, as is our focus here, local equilibrium can also have a meaning for finite systems in close to equilibrium conditions; a main question we ask is that whether the first order non-equilibrium correction to the steady state is of local equilibrium type.

In this paper we study boundary driven Symmetric Inclusion Process (SIP)~\cite{GiardinaRedigVafayi10}, which is a bosonic counter model to the fermionic Symmetric Exclusion Process (SEP)~\cite{Spitzer}. For the boundary driven SEP and also its asymmetric version (ASEP) exact results for the stationary state
were obtained via a matrix formalism \cite{Derrida92, Derrida93, Derrida2007}. Similar approach has not yet been successfully applied to the SIP~\cite{private_giardina}, mainly due to the fact that the particle states in SIP are unbounded in contrast to the bounded states in SEP; i.e. at any site in SIP there can be an arbitrary number of particles, while in SEP the maximum occupancy is one.
In this paper thus we follow a different strategy. We couple the system to two particle reservoirs that are nearly identical, differing by a factor $\varepsilon \ll 1$, in order to keep the system close to equilibrium. There will be, however, some non-reversibility and current flow of particles and as a consequence entropy production of the order of $\varepsilon$. We studied earlier another model of interacting diffusion type  \cite{RedigVafayi11} in related weak coupling settings.

The plan of the paper is as follows. In Section \ref{sip}, we introduce the SIP in contact with particle reservoirs and review its equilibrium properties, and the corresponding thermodynamic potential. Later in Section \ref{sec: irreversible process}, we introduce the irreversible model with slightly perturbed boundary reservoirs and derive the corresponding `external' force on the system.
In Section \ref{st_density_prof}, we present an elementary derivation of the stationarity density profile and the corresponding local equilibrium measure, and in particular their first order expansion.
The result for reversible measure and that of stationary profile were known from \cite{GiardinaRedigVafayi10} and \cite{giardina2013} respectively. Since they are our main starting point, we review them with our choice of model parameters, for the sake of completeness. We  provide a more elementary derivation of stationary profile without explicitly using the concept of duality as in~\cite{giardina2013}.

We proceed in Section~\ref{section_mclennan} to approve the McLennan proposal~\cite{MCLENNAN59, Maes2010} that the first order non-equilibrium correction to the equilibrium measure is the entropy production. The details of the calculation is presented in the Appendix. We then confirm in Section \ref{expansion} that a formal first order expansion calculation in orders of $\varepsilon$ for the non-equilibrium correction to the stationary measure, as expected, yields the McLennan formula. We also indicate a recursive formula, from which one should, in principle, be able to obtain higher order corrections in term of the lower order ones.

Comparing the entropy production and the local equilibrium measure, in Section~\ref{section_mclennan} we find that the first order correction of the stationary measure is a product measure and corresponds exactly to the local equilibrium measure (LEQ). At the end in the Section~\ref{th_limit} we then discuss the LEQ in thermodynamic limit.

\section{Symmetric Inclusion Process (SIP)} \label{sip}
SIP  is a stochastic lattice gas introduced in \cite{GiardinaKurchanRedig07, GiardinaKurchanRedigVafayi09} and further studied in \cite{GiardinaRedigVafayi10, GrosskinskyRedigVafayi11, Grosskinsky13} and a related model in \cite{Bartlomiej} from the condensation point of view. In SIP there is an effective attraction between particles in neighboring sites. We show the state of the system by $\overrightarrow{\eta}=(\eta_1,...,\eta_N)$ where $\eta_i$ is the number of particles at site $i$. We consider here the nearest neighbor interactions, where the transitions happen when a particle jump to its neighboring site, at an exponential time with a rate that depends on the occupation number of the destination site. The process in bulk part of the system is defined via its generator corresponding to a Markov jump process, acting on the core of smooth functions $f:\N^N \to \mathbb{R}$ as observables of the system,
\begin{equation}
\label{eq: L_0}
L_{bulk} f(\overrightarrow{\eta})=\sum_i \eta_i (m+\eta_{i+1}) \left( f(\overrightarrow{\eta}^{i,i+1}) -f(\overrightarrow{\eta}) \right)+
\sum_i \eta_{i+1} (m+\eta_{i}) \left( f(\overrightarrow{\eta}^{i+1,i}) -f(\overrightarrow{\eta}) \right).
\end{equation}
Here $\overrightarrow{\eta}^{i,i+1}$ denotes the configuration obtained from $\overrightarrow{\eta}$ after a particle jumps from site $i$ to site $i+1$. The rate of such transition is therefore $\eta_i (m+\eta_{i+1})$. $m$ is the parameter in the model and it effectively controls the strength of diffusion in the system. 
The two sums in the generator $L_{bulk}$ correspond to jumps to the right and to the left correspondingly.
We also consider two particle reservoirs, one at each boundary, such that particles 
can be put or removed at sites $1$ and $N$ with rates specific of the reservoirs. The generators corresponding to the reservoirs can be written as,
\begin{equation}
\label{eq: left bdr generator}
B_1 f(\overrightarrow{\eta})= b_1(m+\eta_1)  \left( f(\overrightarrow{\eta}^{1+}) -f(\overrightarrow{\eta}) \right) + d_1 \eta_1 \left( f(\overrightarrow{\eta}^{1-}) -f(\overrightarrow{\eta}) \right),
\end{equation}
and
\begin{equation}
\label{eq: right bdr generator}
B_N f(\overrightarrow{\eta})= b_N(m+\eta_N)  \left( f(\overrightarrow{\eta}^{N+}) -f(\overrightarrow{\eta}) \right) + d_N \eta_N \left( f(\overrightarrow{\eta}^{N-}) -f(\overrightarrow{\eta}) \right),
\end{equation}
for the left and right boundaries, respectively. Here $\overrightarrow{\eta}^{i+}$ and $\overrightarrow{\eta}^{i-}$ are the configurations where a particle is added or removed at the site $i$  with rates $b_i$ and $d_i$, respectively. 

Therefore the Markovian generator for the whole system is
\begin{equation} \label{generator}
L=L_{bulk}+B_1+B_N.
\end{equation}
In the theory of Markov processes \cite{Liggett2010}, the generator determines the time evolution of the process in the following sense 
\begin{equation*}
\frac{d}{dt}<f(\overrightarrow{\eta_t})> = <L\, f(\overrightarrow{\eta_t})>,
\end{equation*}
for any observable $f$ of the state $\overrightarrow{\eta_t}$. Here the symbol $<\,>$ indicates the average in the process.
A stationary measure $\nu(\overrightarrow{\eta})$ for the process can be defined as a measure that satisfies
\begin{equation*}
\int L\, f(\overrightarrow{\eta_t}) d\nu(\overrightarrow{\eta}) = 0,
\end{equation*}
for all functions $f$.
The generator formalism is equivalent to the master equation for the evolution of probability measure of the system,
\begin{equation}
\frac{d}{ dt} \nu(\overrightarrow{\eta}) = L^* \nu(\overrightarrow{\eta})=\sum \lambda(\overrightarrow{\eta}', \overrightarrow{\eta}) \nu(\overrightarrow{\eta}')  - \lambda(\overrightarrow{\eta}, \overrightarrow{\eta}') \nu(\overrightarrow{\eta}),
\end{equation}
where the $L^*$ is the adjoint generator and transition rates $\lambda(\overrightarrow{\eta}', \overrightarrow{\eta})$ can be read from the generator expression. For the states,
\begin{equation*}
x=\overrightarrow{\eta}=(\eta_1,\eta_2,\cdots,\eta_N),
\quad y=\overrightarrow{\eta}'=(\eta_1',\eta_2',\cdots,\eta_N'),
\end{equation*}
they can be verified to be
\begin{align}
\label{eq: rate in general}
 \lambda(x, y) =& b_1(m+\eta_1) \delta_{x^{1+},y}+d_1 \eta_1 \delta_{x^{1-},y}+ b_N(m+\eta_N) \delta_{x^{N+},y}+d_N \eta_N \delta_{x^{N-},y}\notag  \\
 &+ \sum_i \eta_i (m+\eta_{i+1}) \delta_{x^{i,i+1},y}+\eta_{i+1} (m+\eta_{i}) \delta_{x^{i+1,i},y}.
\end{align}

\subsection{Reversible Stationary Measure}
We first calculate the stationary measure of the SIP with closed boundaries and parameter $m$. Consider two state
\begin{equation*}
x=\overrightarrow{\eta}=(\eta_1,\eta_2,\cdots,\eta_N),\quad y=\overrightarrow{\eta}^{i,i+1}=(\eta_1,\cdots,\eta_{i-1},\eta_i-1,\eta_{i+1}+1,\cdots,\eta_N).
\end{equation*}  
By~\eqref{eq: rate in general}, we have
\begin{equation*}
\lambda(x,y)=\eta_i(m+\eta_{i+1}),\quad\lambda(y,x)=(\eta_{i+1}+1)(m+\eta_i-1).
\end{equation*}
As a result,
\begin{equation}
\label{eq: ratio of rate}
\frac{\lambda(x,y)}{\lambda(y,x)}=\frac{\eta_i(m+\eta_{i+1})}{(\eta_{i+1}+1)(m+\eta_i-1)}.
\end{equation}

The process satisfies the condition of detailed balance, and has product invariant measures 
\begin{equation} 
\label{productm}
\nu(\overrightarrow{\eta}) = \prod_{i=1}^N \gamma(\eta_i),
\end{equation}
where the marginals $\gamma$ can be obtained via detailed balance,
\begin{equation} 
\label{det_bal}
\nu(x) \lambda(x,y) = \nu(y) \lambda(y,x).
\end{equation}
Together with~\eqref{eq: ratio of rate} results in
\begin{equation*}
\gamma(\eta_i)\gamma(\eta_{i+1}) \eta_i(m+\eta_{i+1})= \gamma(\eta_i-1)\gamma(\eta_{i+1}+1)(\eta_{i+1}+1)(m+\eta_i-1),
\end{equation*}
or equivalently,
\begin{equation*}
\frac{\gamma(\eta_i) \eta_i}{\gamma(\eta_i-1)(m+\eta_i-1)}= \frac{\gamma(\eta_{i+1}+1)(\eta_{i+1}+1)}{\gamma(\eta_{i+1})(m+\eta_{i+1})}.
\end{equation*}
Since this equation has to be valid for all values of $\eta_i$ and $\eta_{i+1}$, we conclude that 
the r.h.s and the l.h.s are equal to a constant $c$. This suggest a recursive formula for $\gamma$,
\begin{equation*}
\gamma(n+1)=  \frac{(m+n)c}{n}   \gamma(n).
\end{equation*}
It is convenient to write the general solution \cite{GiardinaRedigVafayi10} in terms of $\Gamma$ functions,
\begin{equation} \label{stat_m}
\gamma(n)=  \frac{\theta^n\,\, \Gamma(m+n)}{Z_\theta\, n!\,\,   \Gamma(m)},\quad\text{with}~\quad  \Gamma(s):= \int_0 ^{\infty} x^{s-1} e^{-x} dx,
\end{equation}
where $\theta$ is a parameter, determining the average density of particles in the system, and therefore can be thought of as a chemical potential.
Here $Z_\theta$ is a normalization constant given by
\begin{equation} \label{zedt}
 Z_\theta := \sum \frac{\theta^n\,\, \Gamma(m+n)}{n!\,\,   \Gamma(m)}
 = \frac{1}{(1-\theta)^m}.
\end{equation} 

In the process with open boundaries, on the other hand, equilibrium  corresponds to the case $\theta_0=\frac{b_1}{d_1}=\frac{b_N}{d_N}$ for the particle reservoirs. This has the interpretation that the chemical potential at both boundaries are equal to $\theta_0$. In order to simplify the formulas, we assume without loss of generality that $b_1=b_N=b, d_1=d_N=d$ and hence $\theta_0=\frac{b}{d}$, i.e. we consider two identical particle reservoirs. 
These two possible scenarios are equivalent in equilibrium, where kinetic effects (e.g. absolute value of $b_i$ and $d_i$) do not play a role.

In this case too we have the same reversible stationary measure. The value of $\theta$ in~\eqref{stat_m} can be obtained via considering the transitions at one boundary. For instance, we consider two states differing at the site $1$,
\begin{equation*}
x=\overrightarrow{\eta}=(\eta_1,\cdots,\eta_N),\quad y=x^{1+}=(\eta_1+1,\cdots,\eta_N),
\end{equation*}
and with transition rates,
\begin{align}
&\lambda(x,y)=b_1(m+\eta_1),\quad \lambda(y,x)=d_1(\eta_1+1).
\end{align}
Combining this with the condition of detailed balance in ~\eqref{det_bal} gives that in equilibrium $\theta=\theta_0=\frac{b}{d}$. 
\begin{remark}
Note that this calculation is valid for any arbitrary number of particle reservoirs coupled to the system at different sites, in particular for only one particle reservoir. The system would need to be in contact with at least one reservoir to have the \emph{canonical} reversible measure of \eqref{stat_m}. An isolated system will have the \emph{micro-canonical} reversible measure which is the measure \eqref{stat_m} conditioned on having a \emph{fixed} total number of particles $N_p$, i.e. restricted to the hyperplane $\sum \eta_i = N_p$.
 In the canonical measure, for the system to have a finite density of particles we need $0 < \theta <1$ in~\eqref{stat_m}. This corresponds to having a bigger death rate than birth rate from the particle reservoir, i.e. $b<d$.
\end{remark}
\begin{remark}
The choice of $d=b+m$ for the transition rates at boundaries is somehow special. Looking at the form of the generator for the bulk part of the system, this choice corresponds to introducing two \emph{extra-boundary} sites with indices's $0,N+1$ and freezing the number of particle at these sites to $\eta_0=\eta_{N+1}=b$. Particles from neighboring sites can still jump to and back from these \emph{imaginary} boundary sites, as if they are annihilated or created such that the number of particles at the extra-boundary site stays fixed. In this sense, that is a natural choice for the boundary rates to be made, however, the process with general $d \neq b+m$ is quite possible and also well defined.
\end{remark}

\subsection{Thermodynamic Potential}
In analogy with thermodynamics, it is useful to define a thermodynamic potential $U(x)$ such that in equilibrium
\begin{equation*}
 \nu(x) \propto e^{-U(x)},
\end{equation*}
where the proportionality constant is independent of the state $x$. We choose here, instead, to absorb the proportionality constant in $U$ and write an equality
\begin{equation}
 U(x) = -\log \nu(x).
\end{equation}
Rewriting the detailed balance condition in terms of $U$
\begin{equation}
\label{eq: condition}
\frac{\lambda(x,y)}{\lambda(y,x)}=e^{U(x)-U(y)},
\end{equation} 
with the l.h.s. given in \eqref{eq: ratio of rate}.

Since we have product stationary measures, from~\eqref{productm}, we conclude that $U$ is a sum of single-site potentials $V$
\begin{equation*}
U(x)=U(\overrightarrow{\eta})=\sum_{i=1}^N V(\eta_i),
\end{equation*}
and 
\begin{align}
\label{eq: single-site potential}
V(n) & =-\log \gamma(n) \notag\\
& =-n\log \theta + m \log (1-\theta) + \log(n!) - \log \frac{\Gamma(m+n)}{\Gamma(m)}.
\end{align}

As a side-check, since in the bulk dynamics, the rates $\lambda(x,y)$ are non-zero only when the two states $x,y$ above differ at only two places, namely at sites $i$ and $i+1$. A direct calculation using $\Gamma(z+1)=z\, \Gamma(z)$ shows that
\[
U(\overrightarrow{x})-U(\overrightarrow{y})=\log\frac{\eta_i}{m+\eta_i-1}-\log\frac{\eta_{i+1}+1}{m+\eta_{i+1}}.
\]
Similar calculations can be done for the boundary driven transitions at sites 1 and N. Thus equations~\eqref{eq: ratio of rate} and \eqref{eq: condition} are satisfied.

\subsection{Irreversible Process}
\label{sec: irreversible process}
To obtain an irreversible process, we consider a small perturbation of the condition $\theta_0=\frac{b_1}{d_1}=\frac{b_N}{d_N}$ to achieve a system coupled to two particle reservoirs with slightly different chemical potentials. This can be done, for instance, by taking
\begin{equation}
\label{eq: non-equilibrium perturbation}
b_1=b+\varepsilon b, d_1=d;\quad b_N=b-\varepsilon b, d_N=d.
\end{equation}
This means that we slightly perturb the system out of equilibrium, by increasing the birth-rate at the left boundary while decreasing it at the right boundary \footnote{One might consider also the rates $b_1=b+\varepsilon b, d_1=d;\quad b_N=b, d_N=d$ for the irreversible process, but such choice results in a perturbed process which is not absolutely continuous with respect to the original unperturbed process. As a result, the McLennan calculation which rely on the Girsanov formula, is not well-defined and possible.}.

Following the definitions in \cite{Maes2010} of \emph{local detailed balance}, which is a particular perturbation of the transition rates in the process such that
\begin{equation}
\frac{\lambda(x,y)}{\lambda(y,x)}=e^{U(x)-U(y)+F_\varepsilon(x,y)},
\end{equation}
we find what is equivalent of an external force, $F_\varepsilon(x,y)$, corresponding to the irreversible boundaries.  Let us first consider these two states and their corresponding transition rates
\begin{align}
&x=\overrightarrow{\eta}=(\eta_1,\cdots,\eta_N),\quad x^{1+}=(\eta_1+1,\cdots,\eta_N),\label{eq: leftbdr states}\\
&\lambda(x,x^{1+})=b_1(m+\eta_1),\quad \lambda(x^{1+},x)=d_1(\eta_1+1)\label{eq: left bdr rate},
\end{align}
Hence,
\begin{equation}
\label{eq: ratio of rate at left bdr}
\frac{\lambda(x,x^{1+})}{\lambda(x^{1+},x)}=\frac{b_1(m+\eta_1)}{d_1(\eta_1+1)}.
\end{equation}
The external force $F_\varepsilon(x,y)$ must satisfy
\begin{equation}
\label{eq: condition left bdr}
\frac{\lambda(x,x^{1+})}{\lambda(x^{1+},x)}=e^{U(x)-U(y)+F_\varepsilon(x,x^{1+})}=\frac{\lambda_0(x,x^{1+})}{\lambda_0(x^{1+},x)}e^{F_\varepsilon(x,x^{1+})}.
\end{equation}
From~\eqref{eq: ratio of rate at left bdr} and~\eqref{eq: condition left bdr}, it implies that
\begin{equation}
\label{eq: F at left bdr} F_\varepsilon(x,y)=\log(1+\varepsilon).
\end{equation}
Defining $F_\varepsilon(x,y)=\varepsilon F_1(x,x^{1+})+O(\varepsilon^2)$, it follows that \[F_1(x,x^{1+})=1.\]
Similarly for the other cases 
\begin{enumerate}[(1)]
\item $x=\overrightarrow{\eta}=(\eta_1,\cdots,\eta_N),\quad y=x^{1-}=(\eta_1-1,\cdots,\eta_N),\quad F_1(x,x^{1-})=-1;$
\item $x=\overrightarrow{\eta}=(\eta_1,\cdots,\eta_N),\quad y=x^{N+}=(\eta_1,\cdots,\eta_N+1),\quad F_1(x,x^{N+})=-1;$
\item $x=\overrightarrow{\eta}=(\eta_1,\cdots,\eta_N),\quad y=x^{N-}=(\eta_1,\cdots,\eta_N+1),\quad F_1(x,x^{N-})=1.$
\end{enumerate}

\begin{remark}
The external force $F$ is anti-symmetric and non-reversible, i.e., $F(x,y)=-F(y,x)$ and satisfies the following property: for at least one set of states $x_1,x_2,\ldots, x_{n-1},x_n=x_1$
\[
\phi_F(x_1,\ldots,x_n)\colonequals F(x_1,x_2)+F(x_2,x_3)+\ldots+F(x_{n-1},x_n)\neq 0.
\]
Note that $G(x,y)\colonequals U(x)-U(y)$ while anti-symmetric, its not irreversible: $\phi_G(x_1,\ldots,x_n)=0$.
\end{remark}

\section{Stationary density profile and the local equilibrium measure} \label{st_density_prof}
Here we consider the general non-equilibrium process with generator \eqref{generator}, where the corresponding rates in particle reservoirs are $b_1, d_1, b_N$ and $d_N$.
We define the average density of particles at the site $i$ to be
\begin{equation}
 \rho_i=<\eta_i>_{\nu_s},
\end{equation}
where the average is taken according to the stationary measure $\nu_s$ satisfying
the stationarity condition
\begin{equation*}
\int L\, f(\overrightarrow{\eta_t}) d\nu_s(\overrightarrow{\eta}) = 0,
\end{equation*}
for all functions $f$.
Setting $f_i(\overrightarrow{\eta_t})=\eta_i$, a direct calculations shows that for all $2 \leq i \leq N-1$
\begin{equation*}
Lf_i(\overrightarrow{\eta_t})=m\left(\eta_{i-1}+\eta_{i+1}-2\eta_i\right),
\end{equation*}
and for the boundaries,
\begin{align*}
& Lf_1(\overrightarrow{\eta_t})=b_1m+(b_1-d_1-m)\eta_1+m\eta_2, \\
& Lf_N(\overrightarrow{\eta_t})=b_Nm+(b_N-d_N-m)\eta_N+m\eta_{N-1}.
\end{align*}
These, in combination with the stationarity condition give rise to
\begin{align*}
&  \rho_{i-1}+\rho_{i+1}-2\rho_i=0, \\
&  b_1m+(b_1-d_1-m)\rho_1+m\rho_2=0, \\
& b_Nm+(b_N-d_N-m)\rho_N+m\rho_{N-1}=0.
\end{align*}
One way to solve these set of equations is to use an anzats $\rho_i=\alpha+\beta i$ with two unknown parameters $\alpha$ and $\beta$. This anzats automatically satisfies the first equation. From the other two equations we obtain,
\begin{align}
 & \alpha=\frac{b_1\left( d_N-b_N\right) \,m\,N+\left( b_N+b_1\right) \,{m}^{2}+b_N\left( b_1-d_1\right) \,m}{(b_N-d_N)(b_1-d_1)N+\left( d_N+d_1-b_N-b_1\right) \,m+\left( b_1-d_1\right)(d_N-b_N) },\label{eq: alpha} \\
 &
 \beta=\frac{\left( b_N\,d_1-b_1\,d_N\right) \,m}{(b_N-d_N)(b_1-d_1)N+\left( d_N+d_1-b_N-b_1\right) \,m+\left( b_1-d_1\right)(d_N-b_N)}\label{eq: beta}.
\end{align}
\begin{remark}
 This formula is in accordance with the result in \cite{giardina2013}. While here we didn't explicitly used the \emph{duality} concept. However, duality and symmetries are the underlying reasons why such a calculation as presented here is possible, i.e. that we get a set of equations for first moments that do not depend on the higher order moments, which are actually more difficult to calculate.
\end{remark}
\begin{remark}
 Acting the generator on the $f_i(\overrightarrow{\eta_t})=\eta_i$ and equating the result to the discrete gradient of the quantity $J_i=\eta_{i+1}-\eta_i$ shows that $J_i$ is the instantaneous particle current on the bond $\{i,i+1\}$ in the system. Its expectation, $J$, in the stationary state is then $J=<J_i>_{\nu_s}=\beta$.
\end{remark}

\subsection{Local-equilibrium measure}
For every general density profile $\rho_i$ we can associate a corresponding $\theta-$profile $\theta_i$, using the equilibrium relation \eqref{eqdensity}
\begin{equation} 
  \rho_i = \frac{m \theta_i}{1-\theta_i},
\end{equation}
despite the fact that the equilibrium only corresponds to a constant density profile.
This suggests a corresponding local-equilibrium measure (LEQ), which, similar to equilibrium measure, is a product measure and defined as,
\begin{equation} \label{productlocalm}
\nu_{LEQ}(\overrightarrow{\eta}) = \prod_{i=1}^N \gamma_{\theta_i}(\eta_i),
\end{equation}
where as in the equilibrium reversible measure \eqref{productm} the marginals $\gamma_{\theta_i}(\eta_i)$ given as
\begin{equation}
\gamma_{\theta_i}(n)=  \frac{\theta_i^n\,\, \Gamma(m+n)}{Z_{\theta_i}\, n!\,\,   \Gamma(m)}.   
\end{equation}
Since there is a one-to-one correspondence between $\rho_i$ and $\theta_i$ we can freely index the local equilibrium measure by either a $\rho-$profile or a $\theta-$profile.

\subsection{Weakly Non-equilibrium case}
In the weakly non-equilibrium case, with rates given as in equation \eqref{eq: non-equilibrium perturbation}, the coefficients $\alpha,\beta$ in~\eqref{eq: alpha}-\eqref{eq: beta} of density profile are simplified to
\begin{align*}
 & \alpha=\frac{(bd-b^2+bd\varepsilon+b^2\varepsilon^2)mN+2bm^2+(b^2-bd+bd\varepsilon-b^2\varepsilon^2)m}{((b-d)^2-b^2\varepsilon^2)N+(2d-2b)m+(b^2\varepsilon^2-(b-d)^2)},
 \\
  &
 \beta=\frac{-2bdm\varepsilon}{((b-d)^2-b^2\varepsilon^2)N+(2d-2b)m+(b^2\varepsilon^2-(b-d)^2)}.
\end{align*}
Now expanding the density $\rho_i$ up to the first order in $\varepsilon$ gives
\begin{equation}
\label{eq: rho_i}
\rho_i=\alpha(\varepsilon)+\beta(\varepsilon) i+O(\varepsilon^2),
\end{equation}
where
\begin{align}
 & \alpha(\varepsilon)=\frac{b\,m}{d-b}+\frac{b\, d\,m\,(N+1)\, \varepsilon}{(d-b)^2 \,N+2\,(d-b)\,m-(d-b)^2},\label{eq: alpha_varep}
 \\
  &
 \beta(\varepsilon)=-\frac{2\,b\,d\,m\, \varepsilon}{(d-b)^2 \,N+2\,(d-b)
  \,m-(d-b)^2}\label{eq: beta_varep}.
\end{align}
As expected, in the case $\varepsilon=0$ we get back to the equilibrium, and obtain that $\rho_i=\frac{b\,m}{d-b}= \frac{m \theta_0}{1-\theta_0}$.

\subsection{$\varepsilon$ dependence of $\theta$ and the corresponding LEQ measure}
The LEQ measure depends explicitly on $\theta$ and the relation between $\theta$ and $\rho$ is non-linear, therefore we need to first find the appropriate expansion coefficients for $\theta$ before proceeding to do the first order expansion for LEQ measure.
Focusing on a \emph{single-site} density $\rho$ we have the corresponding $\theta$ value from \eqref{eqdensity}
\begin{equation} \label{non-l}
  \theta = \frac{\rho}{m+\rho}.
\end{equation}
Now the linear expansion 
\begin{equation*} 
  \rho=\rho^{(0)}+\rho^{(1)} \varepsilon + O(\varepsilon^2),
\end{equation*}
and 
\begin{equation*} 
  \theta=\theta^{(0)}+\theta^{(1)} \varepsilon + O(\varepsilon^2),
\end{equation*}
gives the following relations
\begin{align}
 \theta^{(0)} =& \frac{\rho^{(0)}}{m+\rho^{(0)}},\label{eq: theta0} \\
 \theta^{(1)} =& \frac{m \rho^{(1)}}{(m+\rho^{(0)})^2}.\label{eq: theta1}
\end{align}
Notice that the superscripts $(0),(1)$ does not indicate the sites index, but they show the expansion order.

The LEQ measure \eqref{productlocalm} contains terms with $\theta^n$, which for them we have \footnote{Neglecting the normalization constant $Z_{\theta}$, it can be shown that its contribution to the expansion is zero up to the first order in $\varepsilon$.}
\begin{equation} 
  \theta^n=\left( \theta^{(0)}\right)^n \left(1+n\varepsilon \frac{\theta^{(1)}}{\theta^{(0)}}  \right) + O(\varepsilon^2).
\end{equation}
Now we can express the LEQ \eqref{productlocalm} corresponding to the stationary density profile in terms of equilibrium measure \eqref{productm} up to first order in $\varepsilon$ using~\eqref{eq: theta0}-\eqref{eq: theta1},
\begin{align} \label{productlocalm_eps}
\nu_{LEQ}(\overrightarrow{\eta}) =& \nu_{EQ}(\overrightarrow{\eta})\left(1+\varepsilon \sum_{i=1}^N \frac{\theta_i^{(1)}}{\theta_i^{(0)}}\eta_i  \right) + O(\varepsilon^2) \notag\\
=& \nu_{EQ}(\overrightarrow{\eta})\left(1+\varepsilon \sum_{i=1}^N \frac{m \rho_i^{(1)}}{\rho_i^{(0)}(m+\rho_i^{(0)})}\eta_i  \right) + O(\varepsilon^2),
\end{align}
or equivalently
\begin{align} \label{productlocalm_epsexp}
\nu_{LEQ}(\overrightarrow{\eta})=& \nu_{EQ}(\overrightarrow{\eta})\exp \left( \varepsilon \sum_{i=1}^N \frac{m \rho_i^{(1)}}{\rho_i^{(0)}(m+\rho_i^{(0)})}\eta_i  \right) + O(\varepsilon^2).
\end{align}
From~\eqref{eq: rho_i},~\eqref{eq: alpha_varep} and~\eqref{eq: beta_varep}, we have
\begin{align}
  \rho_i^{(0)}=&\frac{b\,m}{d-b}, \\
  \rho_i^{(1)}=& \frac{b\,d\,m\,(N+1)}{(d-b)^2 \,N+2\,(d-b)
  \,m-(d-b)^2} \notag\\ &- \frac{2\,b\,d\,m \,i}{(d-b)^2 \,N+2\,(d-b)
  \,m-(d-b)^2},
\end{align}
and therefore
\begin{align} \label{productlocalm_epsexp1}
\nu_{LEQ}(\overrightarrow{\eta})=& \nu_{EQ}(\overrightarrow{\eta})\exp \left( \varepsilon \sum_{i=1}^N \frac{N+1-2 \,i}{N-1+
  \frac{2m}{d-b}} \eta_i  \right) + O(\varepsilon^2).
\end{align}
\begin{remark}
 In the special case that $d=b+m$ we obtain
\begin{align*}
  \rho_i^{(0)}=&b, \\
  \rho_i^{(1)}=& \frac{b\,d}{m}\left(1 - \frac{2\,i}{N+1} \right),
\end{align*}
and as a result
\begin{align} \label{productlocalm_epsexp2}
\nu_{LEQ}(\overrightarrow{\eta})=& \nu_{EQ}(\overrightarrow{\eta})\exp \left( \varepsilon \sum_{i=1}^N \left(1 - \frac{2\,i}{N+1} \right) \eta_i  \right) + O(\varepsilon^2).
\end{align}
\end{remark}

\section{First order expansion: McLennan formula} \label{section_mclennan}
In~\cite{MCLENNAN59}, the author introduced the following formula to approximate the stationary density $\rho$ of an open mechanical system away but close to equilibrium,
\begin{equation}
\label{eq: McL formula}
\rho(x)\approx Z^{-1}\exp(-U(x)+W(x)),
\end{equation}
where $Z^{-1}\exp(-U(x))$ is the equilibrium stationary measure, and $W$ is the non-equilibrium correction~\footnote{Note that we have set the inverse temperature $\beta=1$, since we do not study the effect of temperature here.}.
In~\cite{Maes2010,Maes2011}, the authors provided a rigorous interpretation of this formula for Markov jump and diffusion processes. We now recall the result in~\cite{Maes2010} for the case of Markov jump.
Considering a continuous Markov process on a finite state space $\Omega=\{x,y,\ldots\}$. Let $\lambda(x,y)$ be the transition rate between the states $x\rightarrow y$.
The Master equation for the probability $\mu_t(x)$ of state $x$ is given by
\begin{equation}
\frac{d\mu_t(x)}{dt}=\sum_{y\neq x}\{\mu_t(y)\lambda(y,x)-\mu_t(x)\lambda(x,y)\}.
\end{equation}

The equilibrium dynamic (indicated by subscript $0$) fulfills the detail balanced condition\[
\frac{\lambda_0(x,y)}{\lambda_0(y,x)}=\frac{\rho_0(y)}{\rho_0(x)},
\]
where $\rho_0(x)\propto e^{- U(x)}$ for some potential $U$.
In~\cite{Maes2010}, the authors considered a close-to-equilibrium dynamic by replacing the detailed balanced condition by the local detailed balance condition,
\[
\frac{\lambda(x,y)}{\lambda(y,x)}=e^{F(x,y)+U(x)-U(y)},
\]
where $F$ is anti-symmetric and non-conservative force. To parametrize the distance to equilibrium, the authors took $F_\varepsilon(x,y)=\varepsilon F_1(x,y)$. The main result in~\cite{Maes2010} is the following asymptotic formula for the stationary distribution of the close-to equilibrium dynamic 
\begin{equation}
\label{eq: Maes formula}
\rho_\varepsilon(x)=\rho_0(x)\exp\Big\{-\varepsilon \int_0^\infty \langle w_1(x_t)\rangle_x^0\, dt+O(\varepsilon^2)\Big\},
\end{equation}
where $\langle \cdot\rangle_x^0$ is the averaging over the equilibrium reference process started form $x$ and
\begin{equation}
\label{eq: formula for w_1}
w_1(x)=\sum_{y\neq x}\lambda_0(x,y)F_1(x,y)
\end{equation}
is the entropy production rate.
The proof of~\eqref{eq: Maes formula} consists of three main steps. The first step is to connect the distribution $P$ on path-space with driving $F$ with the equilibrium reference distribution $P^0$ starting from the same state $x$ using the Girsanov formula 
\[
d P_x(\omega)=d P_x^0(\omega)\, \,e^{-A(\omega)}.
\]
Where $\omega=(x_t)_{t=0}^T$ is the process trajectory in the time interval $[0,T]$.
Defining $\rho_T^\varepsilon(x)$ as the distribution of process at time $T$ starting from equilibrium distribution $\rho_0(x)$, the second step is to express $\rho_T^\varepsilon(x)$ in terms of the time-antisymmetric part, $S^T_{\text{IRR}}$, of the action $A$,
\[
\rho_T^\varepsilon(x)=\rho_0(x)\langle e^{-S^T_{\text{IRR}}}\rangle_x.
\]
The time-antisymmetric part of action is given by
\[
S^T_{\text{IRR}}(\omega)=A(\theta\omega)-A(\omega),
\]
where $\theta\omega=(x_{T-t})_{t=0}^T$ for any $\omega=(x_t)_{t=0}^T$.

The last step is to calculate $S^T_{\text{IRR}}(\omega)$. For the Markov jump process, it was proven that
\[
\lim_{\varepsilon \to 0} \frac{1}{\varepsilon} \log \langle \exp \left( -S^T_{\text{IRR}}(\omega) \right) \rangle_x= - \int_0^T \langle w_1(x_t)\rangle_x^0 \, dt.
\]
This, together with $\rho_\varepsilon=\lim_{T\rightarrow \infty}\rho_T^\varepsilon$, gives the McLennan formula~\eqref{eq: McL formula}.
It is worth comparing the two formulas~\eqref{eq: McL formula} and~\eqref{eq: Maes formula}: the result in~\cite{Maes2010} provides an explanation for the McLennan formula in the sense that it identifies the correction term $W$ as the transient part of the irreversible entropy production for the process started from state $x$.

We will apply the method in~\cite{Maes2010} as described above to the SIP. The small perturbation has already been introduced in Section~\ref{sec: irreversible process}. We now compute $w_1(x)$. By~\eqref{eq: formula for w_1},
\begin{align}
\label{eq: w1}
w_1(x)&=\sum_{y\neq x}\lambda_0(x,y)F_1(x,y)\notag
\\&=\lambda_0(x,x^{1-})F_1(x,x^{1-})+\lambda_0(x,x^{1+})F_1(x,x^{1+})+\lambda_0(x,x^{N+})F_1(x,x^{N+})+\lambda_0(x,x^{N-})F_1(x,x^{N-}).
\end{align}
By~\eqref{eq: leftbdr states}-\eqref{eq: left bdr rate}, we have
\[
\lambda_0(x,x^{1+})=b(m+\eta_1),\quad \lambda_0(x,x^{1-})=d \eta_1, \lambda_0(x,x^{N+})=b(m+\eta_N),\quad \lambda_0(x,x^{N-})=d\eta_N. 
\]
In Section~~\ref{sec: irreversible process}, we already calculated
\[
F_1(x,x^{1+})=F_1(x,x^{N-})=1,\quad F_1(x,x^{1-})=F_1(x,x^{N+})=-1.
\]
Therefore,
\begin{equation}
\label{eq: w_1(x)}
w_1(x)=b(m+\eta_1)-d\eta_1-b(m+\eta_N)+d\eta_N=(b-d)(\eta_1-\eta_N).
\end{equation}
\begin{remark}
We now check that $\langle w_1(x)\rangle_\nu=0$, i.e. the mean rate of entropy production in \emph{equilibrium} is zero. To see this, note that for the single site occupation average according to measure $\nu(\overrightarrow{\eta})$ we get
\begin{align}
  <\eta_i>_{\nu(\overrightarrow{\eta})}  & = 
  \sum \eta_i \gamma(\eta_i) \notag\\
& =\sum \frac{\eta_i \theta^{\eta_i}\,\, \Gamma(m+\eta_i)}{Z_\theta\, \eta_i!\,\,   \Gamma(m)}
 = \theta \frac{\partial}{ \partial \theta} \log Z_\theta = \frac{m \theta}{1-\theta}, \label{eqdensity}
\end{align}
where we have made use of~\eqref{zedt} in the last step.
\end{remark}
To obtain the McLennan formula, we need to compute the time integral of $\langle w_1(x_t)\rangle_x^0$.
In the appendix we show that
\[
\lim_{T\rightarrow\infty}\int_0^T\langle w_1(\overrightarrow{x}_t)\rangle_{x_0}^0\,dt
=-L^{-1}w_1(x_0)=-\sum_{i=1}^N c_i \eta_i,
\]
where the coefficients $c_i$ given as
\begin{align*}
c_i=A+Bi \quad \text{for all $1 \leq i \leq N$},
\end{align*}
with
\begin{align*}
  A=\frac{N+1}{N-1-\frac{2m}{b-d}},\,\,\, \text{and} \,\,\, 
 B=\frac{-2}{N-1-\frac{2m}{b-d}}.
\end{align*}

This is identical to the local equilibrium measure corresponding to the stationary density profile that we obtained in section \ref{st_density_prof}. I.e. the first order non-equilibrium correction to the steady state is exactly the LEQ.
\subsection{Thermodynamic Limit: $N \to \infty$ } \label{th_limit}
It is interesting to compare SIP with the case of boundary driven symmetric exclusion process studied in \cite{Bahadoran} with boundaries arbitrarily far from each other,  where it is shown that the Gibbs-Shannon entropy of the stationary state measure converges to that of LEQ in the thermodynamic limit as $N \to \infty$.
In other words, it was shown in \cite{Bahadoran}, for a rather general class of models, that local equilibrium is sufficient to describe the leading-order asymptotic of Gibbs-Shannon entropy. The result we obtained in this paper is in a sense more macroscopic, we study finite size systems (fixed $N$) with boundaries which are nearly identical. It is worth mentioning that the proof in  \cite{Bahadoran} does not directly apply to SIP; SIP lacks at least one of the sufficient conditions in the general proof. This is due to the fact that the particle states in SIP and as a result the entropy production rate are unbounded.

We now discuss the thermodynamic limit for SIP. Consider coupling SIP to two particle reservoir that are arbitrarily different, for instance with $\varepsilon$ not being small, and with very large $N$. Now, looking at a section of the system with size $L\ll N$, for instance $x_i^{(L)}=(\eta_i, ...,\eta_{i+L-1})$, the density profile looks similar to a system of size $L$ which is coupled to two particle reservoirs whose chemical potential difference is small and of the order of $\frac{L}{N}\ll 1$. This suggests the following.
Intuitively, one might expect that in the \emph{thermodynamic limit} and with two boundary reservoirs with arbitrary chemical potentials, the first order non-equilibrium correction to the stationary state is again LEQ corresponding to the density profile. This is the property that was shown rigorously for SIP in \cite{redig_opoku} with the help of the probabilistic technique of \emph{coupling}.

\section{Another way to look at it: Dyson expansion} \label{expansion}
We can write the generator as a sum of the reversible part and an external perturbation with strength $\varepsilon$
\[
L=L_0+\varepsilon \Gamma.
\]
Expanding the (yet unknown) stationary measure up to first order of $\varepsilon$ in terms of equilibrium measure
\[
\rho=\rho_0(1+\varepsilon h)+O(\varepsilon^2)
\]
which must satisfy the stationarity condition
\[
L^*\rho=0.
\]
or
\[
0=L^*\rho=(L_0^*+\varepsilon\Gamma^*)(\rho_0(1+\varepsilon h)+O(\varepsilon^2))=L_0^*\rho_0+\varepsilon(L_0^*(\rho_0 h)+\Gamma^*\rho_0)+O(\varepsilon^2).
\]
This implies that
\begin{equation}
\label{eq: eqn for h}
L_0^*(\rho_0h)=-\Gamma^*\rho_0,
\end{equation}
or equivalently
\[
h=-\frac{1}{\rho_0}(L_0^*)^{-1}(\Gamma^*\rho_0).
\]
For the perturbation corresponding to the boundary rates considered in Section~\ref{sec: irreversible process}, we have
\begin{align*}
&\Gamma f(\overrightarrow{\eta})=b(m+\eta_1)(f(\overrightarrow{\eta}^{1+})-f(\overrightarrow{\eta}))-b(m+\eta_N)(f(\overrightarrow{\eta}^{N+})-f(\overrightarrow{\eta})),
\end{align*}
and
\begin{equation*}
\Gamma^*\nu(\overrightarrow{\eta})=b(m+\eta_1-1)\nu(\overrightarrow{\eta}^{1-})-b(m+\eta_1)\nu(\overrightarrow{\eta})-b(m+\eta_N-1)\nu(\overrightarrow{\eta}^{N-})+b(m+\eta_N)\nu(\overrightarrow{\eta}).
\end{equation*}
The equilibrium measure has been computed in~\eqref{stat_m},
\begin{equation*}
\rho_0(\overrightarrow{\eta})=\prod_{i=1}^N\gamma(\eta_i), \quad \gamma(n)=\frac{\theta^n\Gamma(m+n)}{z_\theta n!\Gamma(m)},\quad z\Gamma(z)=\Gamma(z+1).
\end{equation*}
We get
\begin{align}
\label{eq: Gamma rho}
\Gamma^*\rho_0(\overrightarrow{\eta})= & b\bigl[(m+\eta_1-1)\gamma(\eta_1-1)-(m+\eta_1)\gamma(\eta_1)\bigr]\times\prod_{i=2}^N\gamma(\eta_i)\\
& +b\bigl[(m+\eta_N)\gamma(\eta_N)-(m+\eta_N-1)\gamma(\eta_{N}-1)\bigr]\times\prod_{i=1}^{N-1}\gamma(\eta_i). \nonumber
\end{align}
The first term on the r.h.s of~\eqref{eq: Gamma rho} can be transformed
\begin{align*}
(m+\eta_1-1)\gamma(\eta_1-1)&=
\frac{(m+\eta_1-1)\theta^{\eta_1-1}\Gamma(m+\eta_1-1)}{z_\theta (\eta_1-1)!\Gamma(m)}
\\&=\frac{\theta^{\eta_1-1}}{z_\theta\Gamma(m)(\eta_1-1)!}\Gamma(m+\eta_1)
\\&=\frac{1}{\theta}\eta_1\gamma(\eta_1).
\end{align*}
Similarly, we can rewrite the last term on the r.h.s of~\eqref{eq: Gamma rho} as
\begin{equation*}
(m+\eta_N-1)\gamma(\eta_N-1)=\frac{1}{\theta}\eta_N\gamma(\eta_N).
\end{equation*}
Hence
\[
\Gamma^*\rho_0=\left[\frac{b}{\theta}\eta_1-b(m+\eta_1)\right]\rho_0+\left[b(m+\eta_N)-\frac{b}{\theta}\eta_N\right]\rho_0
=(b-d)(\eta_N-\eta_1)\rho_0.
\]
Substituting to~\eqref{eq: eqn for h}, we obtain
\[
L_0^*(h\rho_0)=\rho_0L_0h=-\Gamma^*\rho_0=(b-d)(\eta_1-\eta_N)\rho_0.
\]
Therefore, finally we recover, as expected, the first order correction the same as in the McLennan formula
\begin{equation}
\label{eq: L_0h}
L_0h=(b-d)(\eta_1-\eta_N).
\end{equation}
\begin{remark}
The equality $L_0^*(h\rho_0)=\rho_0L_0h$ can be verified as follows. 
First we define the following notation for the function $f$ and measure $\rho$,
\begin{equation*}
\langle f,\rho \rangle := \int f \, d\rho.
\end{equation*}
Using definition of $L_0^*$ and reversibility of the process with respect to $\rho_0$, we have for every $f$,
\begin{equation*}
\langle f,L_0^*(\rho_0 h)\rangle=\langle L_0f,\rho_0 h\rangle=\langle f,\rho_0 L_0h\rangle.
\end{equation*}
This implies that $L_0^*(h\rho_0)=\rho_0L_0h$. 
\end{remark}

\subsection{Higher order expansions}
Here we mention the possibility of obtaining higher order corrections.
Extending the first order expansion to a $M$th order expansion we have,
\[
\rho=\rho_0(1+\sum_{i=1}^M \varepsilon^i h_i)+O(\varepsilon^{M+1}).
\]
And from stationarity condition, we have
\begin{align*}
 0=L^*\rho=&(L_0^*+\varepsilon\Gamma^*)\rho_0(1+\sum_{i=1}^M \varepsilon^i h_i)+O(\varepsilon^{M+1})
 \\ =&L_0^*\rho_0+\varepsilon(L_0^*(\rho_0 h_1)+\Gamma^*\rho_0)+\sum_{i=2}^M \left(L_0^*(\rho_0 h_i)+\Gamma^*(\rho_0 h_{i-1})\right) \varepsilon^i+O(\varepsilon^{M+1}).
\end{align*}
Equating the coefficients of different powers of $\varepsilon$ to be zero gives the following recursive formulas for $h_i$
\begin{align*}
h_1 = &-\frac{1}{\rho_0}(L_0^*)^{-1}(\Gamma^*\rho_0), \\
h_i=& -\frac{1}{\rho_0}(L_0^*)^{-1}(\Gamma^*\rho_0 h_{i-1}) \quad \text{for all $2 \leq i \leq M$}.
\end{align*}

\section*{Appendix: Detailed calculation of the McLennan formula} \label{appx}
From the theory of continuous-time Markov processes \cite{Liggett2010} and by definition of the generator we have that for all smooth functions $f$
\[
\frac{\partial}{\partial t}\langle f(\overrightarrow{x_t})\rangle_{x_0}^0=\langle L f(\overrightarrow{x}_t)\rangle_{x_0}^0.
\]
The Markov process defines a semigroup
\[
S_t f(x)=\langle f(\overrightarrow{x_t})\rangle_{x}^0.
\]
By definition of the semi-group
\[
S_t=e^{tL}.
\]
Formally, it follows that
\begin{align*}
\lim_{T\rightarrow\infty}\int_0^T\langle f(\overrightarrow{x}_t)\rangle_{x_0}^0\,dt
&=\lim_{T\rightarrow\infty}\int_0^T e^{tL}f(x_0)\,dt
\\&=\lim_{T\rightarrow\infty}L^{-1}e^{tL}f(x_0)\Big|_{t=0}^T
\\&=-L^{-1}f(x_0)+L^{-1}\lim_{T\rightarrow\infty}e^{TL}f(x_0)
\\&=-L^{-1}f(x_0)+L^{-1}\lim_{T\rightarrow\infty}\langle f(\overrightarrow{x_T})\rangle_{x_0}^0.
\end{align*}
For irreducible Markov process, there is a unique equilibrium measure $\nu$ and starting from any state $x$ the dynamic will converge exponentially fast to this equilibrium state. Hence $\lim_{T\rightarrow\infty}\langle f(\overrightarrow{x_T})\rangle_{x_0}^0=\langle f(\overrightarrow{x_T})\rangle_{\nu}$. For computation of the McLennan formula, $f=w_1$ and we know that $\langle w_1(x)\rangle_{\nu}=0$. Therefore,
\[
\lim_{T\rightarrow\infty}\int_0^T\langle w_1(\overrightarrow{x}_t)\rangle_{x_0}^0\,dt
=-L^{-1}w_1(x_0).
\]

\subsection{Calculation of $L^{-1}w_1(x)$}

We need to calculate $\Phi(x)$ defined as
\begin{equation}
\Phi(x):=L^{-1}w_1(x),
\end{equation}
where $L$ is the generator of reversible process \eqref{generator} with $b_1=b_N=b$ and $d_1=d_N=d$, and $w_1(x)$ given in \eqref{eq: w_1(x)}.
Applying both sides of equation with $L$ we obtain
\begin{equation}
L \Phi(x)=w_1(x).
\end{equation}
This has the advantage that we avoid calculating $L^{-1}$ explicitly.
One way to solve this equation is to use an anzats for $\Phi(x)$
\begin{equation}
\Phi(x)=\sum_{i=1}^N c_i \eta_i.
\end{equation}
Now acting $L$ on $\Phi(x)$ gives
\begin{align*}
L\Phi(x)=& \sum_{i=1}^N c_i L \eta_i \\
 =& \sum_{i=2}^{N-1} m c_i (\eta_{i-1}+\eta_{i+1}-2\eta_i)+ \\
& c_1(bm+(b-d-m)\eta_1+m\eta_2)+c_N(bm+(b-d-m)\eta_N+m\eta_{N-1})
\end{align*}
Using summation by parts to make explicit the coefficients of $\eta_i$, we get
\begin{align}
L\Phi(x)=& \sum_{i=3}^{N-2} m \eta_i (c_{i-1}+c_{i+1}-2c_i)+ 
m c_2(\eta_1-2\eta_2)+m c_{N-1}(\eta_{N}-2\eta_{N-1})
\\
&+ m c_3\eta_2 + m c_{N-2} \eta_{N-1}\\
& +c_1(bm+(b-d-m)\eta_1+m\eta_2)+c_N(bm+(b-d-m)\eta_N+m\eta_{N-1}).
\end{align}
This can be simplified to
\begin{align*}
L\Phi(x)=& 
(c_1+c_N)bm+(c_1(b-d-m)+m c_2)\eta_1 +(c_N(b-d-m)+mc_{N-1})\eta_{N}
\\ &+ \sum_{i=2}^{N-1} m \eta_i (c_{i-1}+c_{i+1}-2c_i).
\end{align*}
Equating this to
\begin{equation}
L\Phi(x)=w_1(x)=(b-d)(\eta_1-\eta_N),
\end{equation}
we will need that the coefficients of all powers of $\eta_i$ for all $i$ be equal on both side of the equation. This results in
\begin{equation*}
 c_{i-1}+c_{i+1}-2c_i=0,
\end{equation*}
for all $2 \leq i \leq N-1$, i.e. in the bulk, and the following for the boundary conditions
\begin{align*}
& c_1+c_N = 0, \\
& c_1(b-d-m)+mc_2 =b-d,\\
&c_N(b-d-m)+mc_{N-1}=-(b-d).
\end{align*}
We use again a linear anzats for $c_i$,
\begin{equation*}
 c_i=A+Bi \quad \text{for all $1 \leq i \leq N$},
\end{equation*}
which will automatically satisfy the bulk discrete Laplace equations. We can find the coefficients $A$ and $B$ from the boundary equation which results in
\begin{align*}
  A=\frac{N+1}{N-1-\frac{2m}{b-d}},\,\,\, \text{and} \,\,\, 
 B=\frac{-2}{N-1-\frac{2m}{b-d}}.
\end{align*}

\section*{Acknowledgment}
We would like to thank Christian Maes for useful discussions. KV acknowledges the support of NWO VICI grant 639.033.008.

\bibliography{sip}
\end{document}